\documentclass[12pt]{emulateapj}

\slugcomment{To appear in ApJL}

\shorttitle{IRAS04325C}
\shortauthors{Scholz et al.}

\begin{document}
\bibliographystyle{apj}


\title{IRAS04325+2402C: A very low mass object with an edge-on disk}


\author{Alexander Scholz\altaffilmark{1}, Ray Jayawardhana\altaffilmark{2}, Kenneth Wood\altaffilmark{1}, 
David Lafreni\`ere\altaffilmark{2,5}, Katharina Schreyer\altaffilmark{3}, Ren\'e Doyon\altaffilmark{4}}

\email{as110@st-andrews.ac.uk}

\altaffiltext{1}{SUPA, School of Physics \& Astronomy, University of St. Andrews, North Haugh, St. Andrews, 
KY16 9SS, United Kingdom}
\altaffiltext{2}{Department of Astronomy \& Astrophysics, University of Toronto, 50 St. George Street, Toronto, 
ON M5S 3H4, Canada}
\altaffiltext{3}{Astrophysikalisches Institut und Universit{\"a}ts-Sternwarte Jena, Schillerg{\"a}sschen 2-3,
D-07745 Jena, Germany}
\altaffiltext{4}{D\'epartement de Physique and Observatoire du Mont M\'egantic, Universit\'e de Montr\'eal, C.P. 6128, 
Succursale Centre-Ville, Montreal, QC H3C 3J7, Canada}

\altaffiltext{5}{Visiting Astronomer at the Infrared Telescope Facility, 
which is operated by the University of Hawaii under Cooperative Agreement no. NNX08AE38A 
with the National Aeronautics and Space Administration, Science Mission Directorate, 
Planetary Astronomy Program.}

\begin{abstract}
IRAS04325+2402C is a low luminosity object located near a protostar in Taurus.
We present new spatially-resolved mm observations, near-infrared spectroscopy, and 
Spitzer photometry that improve the constraints on the nature of this source. The 
object is clearly detected in our 1.3\,mm interferometry map, allowing us to 
estimate the mass in a localized disk$+$envelope around it to be in the range of 
0.001 to $0.01\,M_{\odot}$. Thus IRAS04325C is unlikely to accrete significantly more 
mass. The near-infrared spectrum cannot be explained with an extincted photosphere 
alone, but is consistent with a 0.03-0.1$\,M_{\odot}$ central source plus moderate 
veiling, seen in scattered light, confirming the edge-on nature of the disk. Based on 
K-band flux and spectral slope we conclude that a central object mass $\gtrsim 0.1\,M_{\odot}$ 
is unlikely. Our comparison of the full spectral energy distribution, including 
new {\it Spitzer} photometry, with radiative transfer models confirms the high inclination 
of the disk ($\gtrsim 80\arcdeg$), the very low mass of the central source, and the small 
amount of circumstellar material. IRAS04325C is one of the lowest mass objects with a 
resolved edge-on disk known to date, possibly a young brown dwarf, and a likely wide 
companion to a more massive star. With these combined properties, it represents a 
unique case to study the formation and early evolution of very low mass objects.
\end{abstract}


\section{Introduction}
\label{intro}

The origin of brown dwarfs has been one of the most intensely debated subjects in cool stars
research over the past decade. As of today, there is no consensus on the dominant mechanisms 
that determine the low-mass end of the stellar initial mass function. Recent modeling has 
provided predictions for observable properties in the very low mass (VLM) regime, which in 
principle would permit distinguishing between the proposed scenarios. In practice, however, 
it has turned out to be challenging to place firm limits on the formation theory for VLM 
sources \citep[see reviews by][]{2007prpl.conf..459W,2007prpl.conf..443L}.

Of particular relevance for evaluating formation scenarios are the global properties of 
disks and envelopes. The combination of high-resolution imaging and analysis of the spectral 
energy distribution (SED) in the infrared/mm regime is generally acknowledged as the optimum 
source of information for disk/envelope parameters of young stellar sources. As of today, 
constraints on radii and masses of brown dwarf disks are sparse: Recent deep mm observations 
provided first limits on disk masses ($<0.001-0.003\,M_{\odot}$) and outer radii
\citep[10-100\,AU;][]{2006ApJ...645.1498S,2003ApJ...593L..57K}. Only one substellar disk has been 
resolved with high-resolution imaging \citep{2007ApJ...666.1219L}, with a derived radius
of 20-40\,AU.

In this context, the young source IRAS04325+2402C (hereafter: IRAS04325C) in the Taurus
star forming complex may be of particular interest. HST/NICMOS images of the Class I source 
IRAS04325+2402 \citep{1995ApJS..101..117K}, also known as L1535\,IRS, revealed  IRAS04325C as a 
faint companion, at a separation of $8\farcs1$ \citep{1999AJ....118.1784H}\footnote{Note that
the primary itself is likely to be a subarcsecond binary, see \citet{1999AJ....118.1784H}}. The object 
is resolved `into a double-lobed emission separated by a dark lane' \citep{1999AJ....118.1784H}, 
resembling the images of edge-on T Tauri disks. Models including an edge-on disk and an envelope 
are able to reproduce the appearance of the images. From the near-infrared (NIR) photometry, the mass 
of this possible Class I object was estimated to be between 0.02 and 0.06\,$M_{\odot}$, with an upper 
limit of 0.25\,$M_{\odot}$. Based on statistical arguments, \citet{2004A&A...427..651D} find that the 
system is likely to be a physically bound binary  (probability 98\%). 

Thus, IRAS04325C is a potential benchmark object to study substellar formation. In this Letter, 
we present new millimeter interferometry and NIR spectroscopy as well as an analysis of 
the resolved SED for this source, including new {\it Spitzer} photometry, with the aim of 
improving the constraints on the properties of the system, particularly the mass of the central 
object and the disk$+$envelope.

\section{New constraints from resolved observations}
\label{obs}

\subsection{Millimeter interferometry}
\label{mm}

Using the IRAM Plateau de Bure Interferometer \citep{1992A&A...262..624G}, we obtained a 1.3\,mm 
(226.50 GHz) continuum map centered on IRAS04325C, observed on 27 December 2007 (configuration C, 
rms 0.3\,mJy/beam, HPBW $1.55"\times 1.0"$). The final map of 128$\times$128 square pixels with 
$0\farcs53$ pixel size was produced by Fourier transforming the calibrated visibilities 
using natural weighting using the Grenoble Software environment GAG. 

\begin{figure}
\center
\includegraphics[angle=-90,width=6cm]{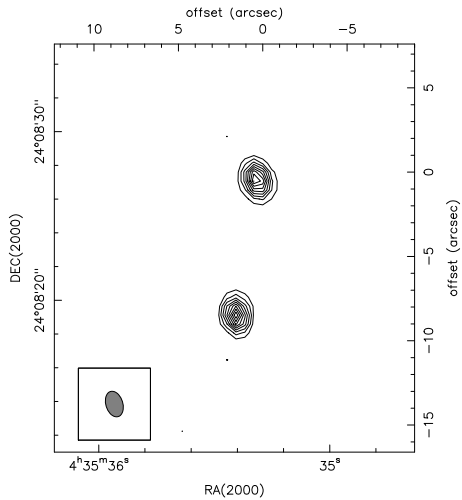}
\caption{Interferometry map at 1.3\,mm wavelength, obtained with the IRAM Plateau de Bure Interferometer. 
The map is centered on our target IRAS04325C; the primary source is seen at (+2",-8"). The inlet in the 
lower left corner shows the beam size. The contours are 10\% to 90\% of the peak value in steps of 10\%. 
\label{f1}}
\end{figure}

The 1.3\,mm image (Fig. \ref{f1}) shows two sources, which are identified with the primary 
(south) and companion (north) object in the HST images from \citet{1999AJ....118.1784H}. We 
estimate fluxes of $10\pm ^3_1$\,mJy for both sources by integrating the pixel intensities. 
The 1.3\,mm flux is  attributed to thermal emission from cold dust. Since the peaks in the
map are strongly centered at the positions of the infrared sources, the emission most likely 
originates from the disk and the localized envelope with a diameter $\lesssim 300\,AU$. Assuming 
that the dust is optically thin, we can obtain an estimate of the total dust mass $M_\mathrm{dust}$ 
from the mm flux $S_\nu$:
\begin{equation}
M_\mathrm{dust} = \frac{S_\nu D^2}{B_\nu(T_\mathrm{D})\kappa_\nu}
\end{equation}
We assume a distance $D$ of 148\,pc for the Taurus star forming region 
\citep{2007ApJ...671..546L}, a gas to dust ratio of 100, and a plausible range of values for 
dust temperature $T_\mathrm{D}$ (10-25\,K) and dust opacity $\kappa_\nu$ (1-3\,cm$^2$g$^{-1}$ at 1.3\,mm), 
as commonly suggested in the literature \cite[see][and references therein]{2006ApJ...645.1498S}. Using 
these parameters, we obtain a total mass (dust and gas) of 1-10\,$M_{\mathrm{Jup}}$ and a most probable 
value of 3\,$M_{\mathrm{Jup}}$ for each of the two sources. This value is comparable to the highest 
disk mass determined for a Class II brown dwarf in Taurus \citep{2006ApJ...645.1498S}.

Single-dish measurements centered on IRAS04325 are available in the literature and can be used 
to put limits on the mass of a possible common envelope surrounding both the primary and the 
secondary. Published fluxes are $\sim 180$\,mJy at 850\,$\mu m$ 
\citep[15" beam,][]{2000ApJ...534..880H,2005ApJ...631.1134A}, 110\,mJy at 1.3\,mm 
\citep[11",][]{2001A&A...365..440M}, $<9.2$\,mJy at 3\,mm \citep[$\sim 11"$,][]{1996ApJ...466..317O}.
Using the same parameters as above and $\kappa_\nu \propto \lambda^{-\beta}$ with $\beta\sim1.5$ 
\citep{2007prpl.conf..767N} yields corresponding masses of 10-30\,$M_{\mathrm{Jup}}$ of distributed 
dust and gas in an area with diameter $\sim 2000$\,AU around the primary IRAS source, including the 
companion. Fluxes integrated over larger areas \citep[500\,mJy for 60" diameter,][]{2001A&A...365..440M} 
clearly indicate the presence of large-scale mm background in that area. Given that the primary in the
system is probably more massive than the companion (see Sect. \ref{nature}), IRAS04325C is unlikely to
be the dominant center of accretion in that area. Thus, the total mass reservoir for IRAS04325C is 
1-10$\,M_{\mathrm{Jup}}$ in the localized disk/envelope plus presumably a small fraction of at 
most 30\,$M_{\mathrm{Jup}}$ in a possible common envelope. 

\subsection{Near-infrared spectroscopy}
\label{spec}

\begin{figure}
\center
\includegraphics[angle=-90,width=8cm]{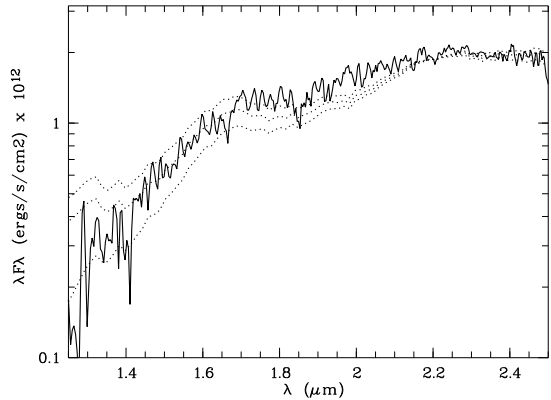}
\caption{IRTF/SpeX spectrum for IRAS04325C (solid line) in comparison with models (dotted lines). All
spectra are scaled to the K-band flux. The models assume $T_{\mathrm{eff}}=3000$\,K and typical veiling. 
The three model spectra in the plot are calculated from three scattering models 
\citep[MRN2, 12, 15 in][]{1993ApJ...414..773K}
\label{f2}}
\end{figure}

We obtained a NIR (1-2.5\,$\mu m$) low-resolution ($R\sim 100$) spectrum for IRAS04325C using SpeX at IRTF
\citep{2003PASP..115..362R}. Reduction, background correction, and extraction was carried 
out using SpeXtools \citep{2004PASP..116..362C}. We corrected for instrument response and telluric 
features using a spectrum of the A0 star HD25175 observed immediately before the target. The final
signal-to-noise ratio ranges from 15 in H- to 30 in K-band. Our spectrum rises steeply from J- to 
K-band and is lacking obvious photospheric features. The spectrum is shown in Fig. \ref{f2} (solid line) 
scaled to the published K-band flux \citep[14.0-14.6\,mag,][]{1999AJ....118.1784H,2008AJ....135.2496C}. In 
the following, we will use the shape of the spectrum and the K-band flux to derive a constraint on 
the mass of the central source.

We compare the observed spectrum with models based on the photospheric templates STARdusty2000
and BDdusty2000 \citep{2001ApJ...556..357A}. In a first attempt, we calculated
a series of models assuming pure extinction, thus neglecting the edge-on nature 
\citep[NIR extinction law from][]{1990ARA&A..28...37M}. While it is possible to reproduce 
the slope of the spectrum using such models, the required combinations of $T_{\mathrm{eff}}$ and 
$A_V$ imply photospheric K-band fluxes that are more than one order of magnitude larger than the 
observed value. Thus, pure extinction cannot account for the NIR SED, confirming that the object 
is seen in scattered light through an edge-on disk.

Thus, instead of extinction, we applied scattering models from \citet{1993ApJ...414..773K}, which have
been confirmed to fit the colours of embedded young stars in Taurus. For a range of scattering properties
and edge-on geometry, \citet{1993ApJ...414..773K} obtain 2.0-4.5\,mag difference between observed and 
photospheric K-band flux, in line with the properties of many known edge-on disks. This implies for 
our target $M_K=3.8\ldots6.9$\,mag. Comparison with the 1\,Myr track by \citet{1998A&A...337..403B} yields $T_{\mathrm{eff}}=2700\ldots3200$\,K and $M=0.03\ldots0.16\,M_{\odot}$.

\begin{figure}
\center
\includegraphics[angle=0,width=8cm]{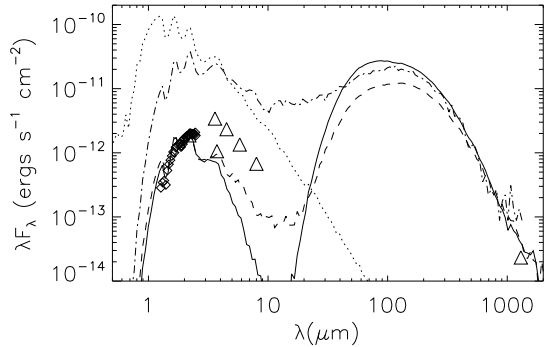}
\caption{Spectral energy distribution for IRAS04325C in comparison with Monte Carlo radiative transfer 
models. Triangles and diamonds are the observed fluxes; Spitzer datapoints at 3.6-8$\,\mu m$ are treated
as upper limits. Solid line: model with ISM grains, dashed line: model with larger grains (see text),
dash-dotted line: model with face-on geometry, dotted line: input photosphere. (All models include 
extinction of $A_V=10$.)
\label{f3}}
\end{figure}

Photospheric spectra for this range of temperatures exhibit deep $H_2O$ absorption features, 
which are not seen in our spectrum. The most likely reason for their absence is veiling, 
as frequently observed for young stellar sources \citep{1996AJ....112.2184G}. We use
the typical veiling for Class-II sources of $r_K=1.76$ and $r_J=0.97$ \citep{1999A&A...352..517F}.
If the K-band flux includes that amount of veiling, the upper limits for temperature and mass are
reduced to $T_{\mathrm{eff}}=3000$\,K and $M=0.1\,M_{\odot}$.

For both scattering and veiling, we fit the values for the photometric bands with a smooth 
function to obtain the wavelength dependence. Veiling is added first, assuming that all excess flux 
originates in regions close to the central object and is thus affected by scattering. Under these 
assumptions, we can fit the observed SED with $T_\mathrm{eff}\sim 3000$\,K, corresponding to  
a mass of $\sim 0.1\,M_{\odot}$ at 1.0\,Myr. In Fig. \ref{f2} we show spectra for three 
plausible scattering models \citep[MRN2, 12, 15 in][]{1993ApJ...414..773K}, all with 
$T_\mathrm{eff}=3000$\,K and veiling, as described above. Stronger veiling, as usually seen for
Class I sources \citep{2005AJ....130.1145D}, would further reduce the photospheric contribution
to the K-band flux and allow for deeper $H_2O$ features, thus lower $T_\mathrm{eff}$ and mass. 

Interpreting the NIR SED for IRAS04325C is challenging and subject to large uncertainties.
We have shown that under reasonable assumptions the observational constraints are well-explained 
by veiled emission from a central object with 0.03-0.1$\,M_{\odot}$ seen through an edge-on disk. 
Higher masses cannot be strictly ruled out, but would require unusual scattering
characteristics. 

\subsection{Spitzer photometry}
\label{spitzer}

IRAS04325 has been observed with Spitzer IRAC/MIPS as part of the Taurus Spitzer Legacy Project (PI: D. Padgett). 
The spatial resolution of MIPS is not sufficient to properly resolve the system; the combined fluxes for the source, 
as given in the Taurus-1 source catalogue, are 1.9\,Jy at 24\,$\mu m$ and $4.7\pm0.9$\,Jy at 70\,$\mu m$. For comparison:
The IRAS mid-infrared fluxes for this object are 2.1\,Jy at 25\,$\mu m$ and 12.86\,Jy at 60\,$\mu m$, again
indicating that long-wavelength data with large apertures include significant extended background emission
(see Sec. \ref{mm}).

In the IRAC images, the object is clearly not point-like and shows two emission peaks coinciding with
the positions of the two components, superimposed onto a bright nebulosity. We carried out resolved
photometry at 3-8\,$\mu m$ by fitting the PSF in apertures with $3\farcs6$ radius centered on the peaks 
of the two components. We estimate the fluxes of the companion to be 4.1, 3.5, 2.6, 1.8\,mJy at wavelengths 
3.6, 4.5, 5.8, and 8.0\,$\mu m$ (uncertainty $\pm 20$\%). Owing to the difficulties in disentangling the 
emission from the sources and from the underlying nebulosity, the true Spitzer fluxes of IRAS04325C are 
likely to be significantly smaller. Comparing with the L-band flux published by \citet{2008AJ....135.2496C} 
indicates that the background contamination may be as much as $\sim 70$\%. The fluxes of the primary source 
in the system are about 6, 7, 10, 11 times higher for IRAC channels 1-4, respectively.

\section{Modeling the SED of IRAS04325C}
\label{model}

We compiled the SED for IRAS04325C from our new observations and the 
literature. In the NIR, we aim to reproduce the K- and L-band fluxes from
\citet{2008AJ....135.2496C}. The Spitzer fluxes are considered to be upper limits, as outlined 
in Sec. \ref{spitzer}. In Fig. \ref{f3} we compare the observed SED with three model SEDs from a 
Monte Carlo radiation transfer code, which has been used successfully to model the SEDs 
of T Tauri stars \citep{2002ApJ...564..887W} and brown dwarfs \citep{2006ApJ...645.1498S}. The models 
use NextGen model atmospheres for the photosphere \citep{2001ApJ...556..357A} and include dust 
destruction close to the central object. Accretion rates are assumed to be negligible for the heating 
of the disk. For more details on the model ingredients, see \citet{2006ApJ...645.1498S}. 
In all three models, we fix the disk radius at 30\,AU, as given by \citet{1999AJ....118.1784H} 
based on the analysis of the HST images, the extinction at $A_V=10$, and the parameters of the 
central object at values consistent with our spectroscopy ($T_\mathrm{eff}=2600$\,K, $M=0.07\,M_{\odot}$).
In addition, the model contains an infalling envelope with a bipolar cavity. 

The models shown with solid and dashed line assume an inclination of $i=80\arcdeg$ and two different
grain size distributions: small ISM-type grains (solid line) and larger grains (dashed line), mimicking 
the effect of dust coagulation. In the latter case, the grain size distribution is a power law with 
an exponential decay for particles with sizes above 50$\,\mu m$ and a formal maximum grain size of 1\,mm,
see \citet{2002ApJ...564..887W}. Both models are able to provide a good match to the available datapoints. 
To fit the mm datapoint we need a total circumstellar mass of $2.2\times 10^{-3}\,M_{\odot}$ for ISM grains 
and $4.6\times 10^{-4}\,M_{\odot}$ for the larger grains. In the second case, the value for the disk/envelope
mass is lower than the constraint given in Sec. \ref{mm} due to the higher dust opacity. Note that the two 
models are highly discrepant in their prediction for the fluxes at  $5-20\,\mu m$ due to the different 
scattering albedo of the two grain types.

For illustration purposes, we also provide the SED for a face-on geometry in dash-dotted lines.
Clearly, the model is not in agreement with the observations. No combination of $T_\mathrm{eff}$ and
$A_V$ is able to fit both the observed fluxes and the near-infrared spectral slope, as discussed in
Sec. \ref{spec}.

In summary, the modeling clearly confirms the plausibility of the results obtained in earlier 
sections. The observed SED can be reproduced assuming that the central source is a very low mass star
or brown dwarf, the inclination is high (i.e. the object is seen mostly in scattered light), and the
circumstellar mass is negligible compared with the object mass. The models also demonstrate 
that more complete wavelength coverage is needed for further analysis of the properties 
of the system.

\section{The nature of IRAS04325C}
\label{nature}

We present new observational constraints on the nature of the potential young brown dwarf IRAS04325C,
a $8\farcs1$ companion to a well-known IRAS source discovered by \citet{1999AJ....118.1784H} in 
HST/NICMOS images. From our NIR spectrum for IRAS04325C, we estimate that the central source 
likely has a mass between 0.03 and 0.1$\,M_{\odot}$. The available constraints on the infrared SED 
are inconsistent with a reddened photospheric spectrum, but they can be understood by assuming that 
the object is seen in scattered light. This is independent confirmation for the edge-on geometry of 
the disk, as seen in the HST images. IRAS04325C is only the second case of an edge-on system in the 
mass range $\lesssim 0.1\,M_{\odot}$ that has been resolved with high-resolution imaging. 

We detect the source with 1.3\,mm interferometry at a flux level of 10\,mJy, corresponding to a mass
of 1-10\,$M_{\mathrm{Jup}}$ in a localized disk/envelope with $R\lesssim 300$\,AU. Since the accretion
reservoir in a larger-scale envelope is probably insignificant, the central source is unlikely to
gain more mass than 0.01\,$M_{\odot}$ due to further accretion. Thus, IRAS04325C will probably become 
either a very low mass star close to the substellar boundary or a brown dwarf. 

The evolutionary stage of IRAS04325C is not well-determined based on the current data.
The primary component of the system has been classified as a Class I source based on the mid-infrared 
spectral slope \citep[$\alpha(2-25)=0.79$,][]{2004A&A...427..651D} and the NIR colours 
\citep{1992ApJ...393..278L}. It is also assumed to drive a CO outflow \citep[e.g.][]{1998ApJ...502..315H}, 
another characteristic for a protostellar source. The companion is clearly embedded in the same cloud core 
and likely physically bound to the primary, thus the two sources can be assumed to be coeval. This would 
make IRAS04325C the lowest mass Class I source known to date. The classification of the primary source, 
however, is partly based on unresolved data, including light from the companion and the surrounding 
nebulosity. Furthermore, the disk of the primary is probably seen close to edge-on 
\citep[$i\sim 80\arcdeg$,][]{2008ApJS..176..184F}, which would suppress the NIR flux and might lead 
to its misidentification as Class I source \citep[see][]{2007prpl.conf..117W}. We estimate a total mass 
of a few Jupiter masses in a localised disk$+$envelope for the primary source, which seems to be too low 
for a Class I object \citep{2005ApJ...631.1134A}. Thus, the Class I nature of the source is questionable. 

The central source of IRAS04325C has probably significantly less mass than the primary component in the system.
In all near/mid-infrared-bands, IRAS04325C is about an order of magnitude fainter than the primary 
source, indicating a mass ratio of $\sim 0.15$. The luminosity of the primary has been estimated to be 
0.7-0.9$L_{\odot}$ \citep{1993ApJ...414..773K,2008ApJS..176..184F}, implying a mass of 
$\sim 0.7\,M_{\odot}$. IRAS04325C also has a much smaller disk than the primary object, as seen 
in the HST images.

With its combined properties -- very low mass central object, edge-on disk, wide companion to a more massive
star -- IRAS04325C is a prime target for follow-up studies and presents us with a unique opportunity 
to study the early evolutionary stages in a critical mass range.

\acknowledgments
This work is based on observations carried out with the IRAM Plateau de Bure Interferometer. IRAM 
is supported by INSU/CNRS (France), MPG (Germany) and IGN (Spain). We would like to thank the staff at 
IRAM, in particular Jan-Martin Winters and Vincent Pietu, for support with observations and data reduction.
The help provided by John Rayner during the SpeX observations is appreciated. Parts of this study 
are based on archival data obtained with the Spitzer Space Telescope, which is operated by the Jet 
Propulsion Laboratory, California Institute of Technology under a contract with NASA. 

\bibliography{aleksbib}

\end{document}